# Visible-light beam size monitors using synchrotron radiation at CESR

S.T. Wang [a,*], D.L. Rubin [a], J. Conway [a], M. Palmer [a], D. Hartill [a], R. Campbell [b], R. Holtzapple [b]

[a]Cornell Laboratory for Accelerator-based Science and Education, Cornell University, Ithaca, New York, 14853

[b]Physics Department, California Polytechnic State University, San Luis Obispo, CA, 93407

*Corresponding author: 218 Wilson Lab, Ithaca, NY 14853. E-mail: sw565@cornell.edu Tel: 607-255-8777. Fax: 607-255-8062

ABSTRACT

A beam profile monitor utilizing visible synchrotron radiation (SR) from a bending magnet has been designed and installed in Cornell Electron-Positron Storage Ring (CESR). The monitor employs a double-slit interferometer to measure both the horizontal and vertical beam sizes over a wide range of beam currents. By varying the separation of the slits, beam sizes ranging from 50 to 500 µm can be measured with a resolution of approximately 5 µm. To measure larger beam size (> 500 µm), direct imaging can be employed by rotating the double slits away from SR beam path. By imaging the $\pi$-polarized component of SR, a small vertical beam size (~70 µm) was measured during an undulator test run in CESR, which was consistent with the interferometer measurement. To measure the bunch length, a beam splitter is inserted to direct a fraction of light into a streak camera setup. This beam size monitor measures the transverse and longitudinal beam sizes simultaneously, which is successfully used for intrabeam scattering studies. Detailed error analysis is discussed.

*Keywords:* Interferometer; Synchrotron radiation; Beam size monitor;

## 1. Introduction

Visible light from synchrotron radiation (SR) has been widely used to measure and monitor the transverse [1] and longitudinal [2, 3] beam sizes in accelerators. Direct imaging is the most common method in which the transverse beam sizes are extracted from Gaussian beam profiles [1, 4]. One important component of this method is the deconvolution of the resolution from the measured beam size. However, as there are many contributions to the optical resolution such as the diffraction limit, depth of field [4], the accuracy with which it can be determined is limited. Thus, when the beam size is comparable to or smaller than the optical resolution, the direct imaging method for measuring the transverse beam size may have a large uncertainty. In the Cornell Electron-Positron Storage Ring – Test Accelerator (CESR-TA) low emittance lattice [5], the anticipated transverse beam sizes at the source point of the visible-light beam size monitor (vBSM) are $\sigma_x$=154 μm and $\sigma_y$=17 μm (Table 1), both well below the calculated optical resolution of 230 μm [4, 6]. Therefore, the direct imaging method using visible light is not suitable for measuring the transverse beam sizes in this lattice.

In order to measure small transverse beam size, many different approaches have been explored. To overcome the diffraction limit, beam size monitors using SR with short wavelength (x-ray) have been developed in many accelerator facilities [7-9]. In CESR-TA, an x-ray beam size monitor was constructed [10-12], which is capable of measuring vertical beam size as small as 10 μm. With the fast-readout electronics, it is also able to perform turn-by-turn and bunch-by-bunch measurements. However, the x-ray monitor was not constructed for simultaneous measurement of both horizontal and vertical beam sizes.

The SR interferometer first applied by Mitsuhashi [13-15] is a well developed instrument which has been constructed in many storage rings and x-ray facilities [16-20]. It was successfully demonstrated to accurately measure transverse beam sizes of less than 300 μm [14]. Two-dimensional interferometers have also been built to measure the horizontal and vertical beam sizes simultaneously [14, 21, 22]. Another approach to measure very small beam size is to image the vertical (π) polarization component of SR light. From the measured peak-to-valley intensity ratio of the vertically polarized beam profile, the vertical beam size can be extracted. This technique has been successfully implemented at the Swiss Light Source and has measured the vertical beam size less than 10 μm [23].

In this paper, we review the configuration of the beam profile monitor developed at CESR for measuring transverse beam sizes. It is demonstrated that beam sizes ranging from 50 to 500 μm can be measured with a resolution of approximately 5 μm using an interferometer. The procedure for calibrating the interferometer is described. We also show that with this vBSM we have the capability to measure the beam sizes in all three dimensions simultaneously, which is necessary for intra-beam scattering studies.

## 2. Experimental methods

### 2.1 Experimental Setup

The beam profile monitors are located in the north area (L3 section) of CESR, diametrically opposite the old CLEO detector. The vBSMs are placed symmetrically at east and west ends of the L3 straight in order to image visible SR from the counter rotating electron and positron beams respectively. Fig. 1a shows the schematic layout of the instrument optics that transports the SR from the CESR accelerator to a radiation-safe experimental hall. The visible SR from a soft bending magnet (140 m bending radius) is reflected by a water-cooled Beryllium (Be) mirror in the vacuum chamber, through a quartz vacuum window and into an optics box (Fig. 1b). The mirror was made by cutting a 1 inch long Be cylinder with a diameter of 0.75 inch diagonally (cutting angle 45°). The dimensions of the polished mirror surface are 22x12.6 mm (HxV) so that the angular acceptance of the Be mirror is 2.5×2.2 mrad (H×V). After passing through an iris with adjustable aperture, the SR photons are incident on a double-slit interferometer. Beyond the slits, and now 6 m from the source in the bending magnet, the light is focused through a lens with 5 m focal length. The iris, the slits, the focusing lens and finally a mirror are assembled in the optics box (Fig. 1b), immediately adjacent to the beam line. With the help of multiple reflecting mirrors, the SR light is transmitted through the wall of the accelerator tunnel to an optics table in the experimental hall, about 27 m from the source point. On the optical table, the SR light passes through a second lens with 1 m focal length, a polarizer, and a 500-nm bandpass filter. We observe the image that emerges from the filter with a CCD camera (Flea3 [24]). A beam splitter can be placed in the path to direct part of the light into a streak camera (Fig. 1a), which is used to measure the bunch length [3].

As evident in Fig. 1a, the maximum angular acceptance is determined by the dimensions of the primary Be mirror, and the minimum by the aperture of the iris. The iris diameter can be varied from 3 to 24 mm, corresponding to angular acceptance of visible light from 0.5 to 3.5 mrad. The depth of field in turn depends on the angular acceptance. With minimum angular acceptance of 0.5 mrad the depth of field is 70 mm. The maximum depth of field is 350 mm. Because the double slits are located inside the accelerator tunnel, adjustment of the slit separation is inconvenient during operation. Therefore, three sets of double-slits with three different slit separations were machined into a single piece of aluminium and mounted on a translation stage and a rotation stage. Three different slit separations were 2.0, 2.5, and 3.0 mm for horizontal beam size measurement, and 3.0, 5.0, and 8.0 for vertical beam size measurement, respectively. The width of all the slits is 0.5 mm. Both rotation and translation stages are remotely controlled. Different sets of slits can be chosen by appropriate setting of the translation stage to measure the transverse beam size over different ranges. The slit sets can also be rotated away from the SR beam in order to image the beam profile directly.

The optical elements on the optics table in the experimental hall are readily accessible and adjustable during storage ring operation. The polarizer can be rotated 90° manually to choose either the σ-polarized or π-polarized component of SR light. The position of the CCD camera can be adjusted to view one or the other of two distinct image planes. The CCD camera can be set to view the source point in order to directly image the beam profile (Fig. 2a) or the interference fringes (Fig. 2b). Alternatively, the camera position can be chosen to image the slits (Fig. 2c), providing a means to determine slit alignment and an opportunity to minimize the imbalance factor (Fig. 2d). All the images were scaled according to the magnification ratio of ~0.2. The images of the CCD camera record the intensity of SR light over a short period of time, which is the exposure time set by the control program. Depending on the light intensity, the exposure time of CCD camera is typically 0.5 to 50 ms, during which the electrons circulate about 200 to 20000 turns (2.56 μs per turn).

CESR is a storage ring for both electron and positron beams with mirror symmetry about the north-south diameter. The two vBSMs are installed symmetrically about the north-south axis, at east and west end of the north area straight, for electron and positron beams, respectively. The lattice functions are the same for both species. The measurements acquired for both electron and positron beams were similar. Unless explicitly indicated, only the results for electron beam size measurements are discussed here.

*2.2 Interferometer theory*

The principles behind the interferometer method for measuring beam size have been discussed in many papers [13, 14]. Here we briefly review the method. The instrument is a wave-front-division-type two-beam SR interferometer using polarized quasi-monochromatic light. When the light intensities at the two slits are the same, the interference pattern at the detector plane can be written as

$$I(x) = I_0 \left[ \text{sinc}(\frac{2\pi a}{\lambda R} x + \phi) \right]^2 \left(1 + |\gamma|\cos(\frac{2\pi D}{\lambda R} x + \psi)\right) \quad (1)$$

where $I_0$ is the light intensity through the slits, $a$ is the slit width, $R$ is the distance between the double slits and the detector, $\lambda$ is the wavelength, and $D$ is the separation of the double slits. $\phi$ and $\psi$ are phase shifts. $\gamma$ denotes the spatial coherence (or visibility). The spatial coherence can evidently be extracted from the measured intensity distribution.

From the van Cittert-Zernicke theorem, the degree of coherence $\gamma$ is the Fourier transform of the source distribution $f(x)$:

$$\gamma = \int dx f(x) \exp(-i\frac{2\pi D}{\lambda L} x) \quad (2)$$

where $L$ is the distance between the source and the double slits. If the beam shape $f(x)$ is a Gaussian profile, its width $\sigma_x$ is related to spatial coherence $\gamma$ according to

$$\sigma_x = \frac{\lambda L}{\pi D} \sqrt{\frac{1}{2} \ln \frac{1}{\gamma}} \quad (3)$$

Therefore, by acquiring and fitting the interference pattern to obtain the spatial coherence $\gamma$, we can measure the beam size $\sigma_x$.

If the beam profile has a uniform distribution, the beam shape function $f(x)$ is

$$f(x) = \frac{1}{2A} \text{ while } -A < x < A; \ = 0 \text{ while } x > A \text{ or } x < -A \quad (4)$$

Then, the visibility $\gamma$ for this uniform distribution is

$$\gamma = \frac{\sin(2\pi DA/L\lambda)}{2\pi DA/L\lambda} \quad (5)$$

By measuring the visibility $\gamma$, the half width $A$ can be determined.

*2.3 Vertical polarization component of SR*

The π-polarization method of measuring the vertical beam size has also been discussed in detail in many references [23, 25, 26]. The underlying principle is to make use of the on–axis minimum of the π-polarized intensity distribution which will be blurred with increasing vertical beam size. Based on a near-field calculation of the SR electromagnetic fields using the retarded potentials, Synchrotron Radiation Workshop (SRW) code simulates the π-polarized intensity distribution in the image plane [27]. The simulation assumes the beam profile has a Gaussian distribution.

Fig. 3a shows a typical π-polarized intensity distribution simulated using SRW. From the vertical profile (Fig. 3b), we can see the valley-to-peak intensity ratio is finite for a finite beam size while the ratio is zero for a point source. By simulating the π-polarized intensity distribution with different vertical beam sizes, we obtain the theoretical curve that relates valley-to-peak intensity ratio to vertical beam size (Fig. 3c). Therefore, by imaging the experimental π-polarized component of SR light and extracting the valley-to-peak intensity ratio, we can infer the vertical beam size.

## 3. Results and analysis

*3.1 Bench test and calibration*

After the vBSM was installed in CESR, all the optical elements (mirrors, iris, and slits) inside and outside the tunnel were aligned. A bench test was developed to align and calibrate the interferometer.

First, a laser was placed at roughly the same distance from the optics box as the source point. A periscope was mounted close to the iris in order to bypass the Be mirror inside the vacuum chamber and to direct the laser light onto the slits. The distance from the laser to the double slits is about 6 m. The mirrors were aligned one by one so that the laser beam was transmitted the length of approximately 27m optical path to the optics table outside the accelerator tunnel.

After the alignment, the laser was replaced with a filament source behind an adjustable width single slit. The CCD camera and other optical elements on the optics table were positioned to image the filament source. The double slits were then inserted to acquire the interference pattern. Fig. 4a shows the interference pattern when the single slit half width was set to 370±10 μm (measured using a microscope). By fitting the integrated intensity profile using Eq. 1 (Fig. 4b), the visibility of 0.44±0.01 was obtained. Assuming a uniform distribution across the single slit, its half width was calculated as 376±6 μm using Eq. 5, consistent with the half width measured using the microscope.

The measurements were repeated several times by varying the width of the single slit. As shown in Fig. 4c, the half widths extracted by the interferometer agree reasonably well with those measured using the microscope and obtained from the direct imaging method, demonstrating the reliability of our optical model and our interpretation of the measurements. Only minor adjustment of the optical elements were required to transition from calibration mode to measurements of transverse beam dimensions using synchrotron radiated light from a circulating bunch of charged particles.

*3.2 Horizontal beam size measurement in low emittance lattice of CESR-TA*

We measured horizontal beam size in the Cesr-TA 2GeV low emittance lattice. A typical interference pattern measured with a set of double slits (*D*=2.0 mm) is shown in Fig. 5a. Fitting the fringes with Eq. 1 yields a visibility $\gamma$=0.52±0.01 (Fig. 5b), which corresponds to a beam width $\sigma_x$=275±4 μm, where we have assumed a Gaussian beam profile (Eq. 3). The visibilities measured with double slits with three different separations (*D*=2.0, 2.5, 3.0 mm) are shown in Fig. 5c. By fitting the visibility as a function of slit separation using Eq. 3, a more accurate horizontal beam size $\sigma_x$=272±1 µm was obtained, which is consistent with the measurement using a single set of slits. Although the second method (scanning slit separation) yields a more accurate measurement of beam size, it is not practical for monitoring a rapidly changing beam size. Therefore, a single set of slits was used to monitor the beam size during the intra-beam scattering (IBS) studies.

*3.2.1 Error analysis*

We consider systematic effects that can contribute to measurement errors. Using neutral density filters to vary the intensity of light reaching the CCD camera, the nonlinearity of the camera response was determined to be negligibly small [15]. The relative intensity on each of the slits is observed by positioning the CCD camera at the image plane of the slits as shown in Fig. 2d. The slits are translated horizontally to minimize any intensity imbalance [13].

Since the SR comes from a horizontal bending magnet, effects due to the curvature of the trajectory and the depth of field make a systematic contribution to the measured horizontal beam size. Including these effects, the visibility $\gamma_h$ is given by a superimposing integration of the van Cittert-Zernike theorem [13]:

$$\gamma_h = \iint \frac{2\sqrt{I_1(\psi)I_2(\psi)}}{I_1(\psi)+I_2(\psi)} f[x-\rho(1-\cos(\psi)),\psi] g(\psi) \exp(-i\frac{2\pi D}{\lambda L}x) d\psi dx \qquad (6)$$

where g($\psi$) is the angular distribution of the SR beam in the horizontal plane as a function of the observation angle $\psi$; $\rho$ is the bending radius; $I_1$ and $I_2$ are the intensities of the light that illuminate the double slits. Assuming the beam has a Gaussian profile

$$f[x-\rho(1-\cos(\psi)),\psi] = \frac{1}{\sqrt{2\pi\sigma_x^2}} \exp\left[-\frac{(x-\rho(1-\cos\psi))^2}{2\sigma_x^2}\right] \qquad (7)$$

the visibility function now becomes

$$\gamma_h = \int \frac{2\sqrt{I_1(\psi)I_2(\psi)}}{I_1(\psi)+I_2(\psi)} g(\psi) \exp\left[-2(\frac{\pi D \sigma_x}{\lambda(L-\rho\sin\psi)})^2 - i\frac{2\pi D\rho(1-\cos\psi)}{\lambda(L-\rho\sin\psi)}\right] d\psi \qquad (8)$$

From Eq. 8, the visibility as a function of the slit separation (*D*) with fixed beam size ($\sigma_x$ =220 µm), and visibility as a function of beam size with fixed slit separation (*D*=2.5 mm), respectively are plotted in Fig. 6. As we can see, the overall effect from depth of field is small, and consistent with the previous simulations [13]. The effect is completely negligible when $\sigma_x$ is greater than 400 µm but is significant when $\sigma_x$ is less than 100 µm. From these simulations, we can determine the systematic correction to the measured beam size due to depth of field.

As noted above, the depth of field depends on the aperture of the iris. The beam size was observed to be constant when the depth of field was varied from 350 to 80 mm for a 250 µm beam, consistent with the results in Fig. 6.

In CESR, we have observed a turn-by-turn beam motion of order 50 µm both horizontally and vertically. This beam jitter can generate a phase shift in the interference pattern that will reduce the visibility and results in an overestimate of the beam size. When the beam is off the nominal orbit, the phase error has been calculated [28] to be

$$\Delta \psi = \frac{2\pi}{\lambda} \frac{D}{L} \Delta x_{beam} \qquad (9)$$

where $\Delta x$ is the beam offset from the orbit. If the beam moves horizontally from $-\Delta x$ to $\Delta x$, the phase shift of the interference pattern will change from $-\Delta \psi$ to $\Delta \psi$. By integrating the interference intensities over this phase shift, we find the average visibility to be

$$\bar{\gamma} = \text{sinc}(\Delta \psi) \gamma \qquad (10)$$

Assuming $\Delta x$=50 µm, we get $\Delta \psi$=0.26 and $\bar{\gamma}$ =0.989$\gamma$. This 1.1% contribution to the visibility corresponds to an error in the measurement of horizontal beam size of 1~2 µm or less than 1%.

As shown in Fig. 5c, the measured visibility has an error bar ±0.01 at fixed slit separation and constant current. This measurement error is due to the fitting error, CCD camera noise, and beam jitter. The error in visibility of $\Delta \gamma$~0.01 results in a measured beam size error $\Delta \sigma_x$ that depends on the slit separation. As shown in Fig. 7, the beam size error is less than 5 µm when the measured visibility is in the range of 0.2-0.8 (the beam size in the range of 50-400 µm for $D$=2.0 mm double slits).

*3.2.2 Horizontal emittance*

The measured horizontal beam size depends on horizontal emittance, energy spread and Twiss parameters $\beta_x$ and $\eta_x$ at the source point. To explore that dependence and in order to extract emittance and energy spread from the size measurements, we created local beta and dispersion bumps to vary only $\beta_x$ and $\eta_x$ respectively at the source point without affecting lattice properties at other regions.

As shown in Fig. 8a, with fixed beam current ~1mA, the horizontal beam size depends as expected on $\beta_x$. A linear fit of $\sigma_x^2$ vs $\beta_x$ with $\eta_x$=0 gives a horizontal emittance $\varepsilon_x$ of 6.0±0.1 nm. Dependence of beam size on horizontal dispersion $\eta_x$ is shown in Fig. 8b. With increasing horizontal dispersion $\eta_x$ at the source point and fixed $\beta_x$ (9.2 m), we observe an increase of horizontal beam size (at fixed beam current 0.6 mA), and find that $\sigma_x^2$ varies linearly with $\eta_x^2$. A linear fit yields both the energy spread and emittance. The measured $\sigma_E/E$=0.08% is very close to the design value of 0.081%. The measured effective horizontal emittance, $\varepsilon_x = \sigma_x^2/\beta_x$, is ~5.7 nm, consistent with the effective horizontal emittance measured by varying $\beta_x$. The normal mode emittance computed from radiation integrals is $\varepsilon_a$~2.6 nm (Table 1). The projected horizontal beam size includes a contribution from the x-z tilt of the bunch at the vBSM source point, which results from the relatively large horizontal dispersion in the RF accelerating cavities ($\eta_x \approx$ 1 m). For a given normal mode emittance, we can estimate the x-z beam tilt $\theta$ at the vBSM source point from Eq. 11.

$$\sigma_x^2 = \beta \varepsilon_a + (\theta \sigma_z)^2 \qquad (11)$$

Given the calculated $\varepsilon_a$ = 2.6 nm and the measured bunch length $\sigma_z$ = 10.5±1.0 mm [29], we find that the x-z beam tilt $\theta$ = 0.017±0.003 rad. The calculated x-z beam tilt at the vBSM source point is 0.015±0.005 rad [30], where the quoted error is a combination of the uncertainties in the relative voltage of east and west RF cavities and the dispersion function in the RF cavities. The measured x-z tilt is in reasonable agreement with the model prediction.

*3.2.3 Intrabeam scattering*

The interferometer CCD camera gain can be adjusted in real time to track beam current, so that we can measure current dependence of beam size. The sensitivity of our camera allows us to measure the horizontal beam size accurately to beam current as low as 0.1 mA ($1.6 \times 10^9$ particles/bunch). We have successfully measured the dependence of horizontal beam size on bunch current while the single bunch current varied from 8 mA to 0.1 mA to study IBS.

Fig. 9 shows a typical measurement of the horizontal beam size as a function of beam current. We see that the IBS simulation (red line) is in reasonable agreement with the experimental data, indicating the horizontal beam size is IBS-dominated. A detailed discussion of IBS can be found elsewhere [29].

*3.2.4 Optical resolution*

For the direct imaging method, the optical resolution ($\sigma_r$) needs to be subtracted from the measured raw beam size ($\sigma_m$) to obtain the corrected transverse beam size ($\sigma^2 = \sigma_m^2 - \sigma_r^2$). Normally, the resolution is calculated based on the diffraction, depth of field, curvature effects, and etc. [4, 6]. Since the horizontal beam size ($\sigma_x$) can be measured accurately using the interferometer, we can obtain the horizontal optical resolution by subtracting the accurate horizontal beam size from the measured beam size ($\sigma_r^2 = \sigma_m^2 - \sigma_x^2$).

For example, at a fixed beam current of 0.6 mA, the horizontal beam size first measured by the interferometer was $\sigma_x$=244.6 μm. We rotated the double slits out of the beam path to image the beam profile directly. With the magnification factor of approximately 0.2 and the CCD camera pixel size of 4.4 μm, the measured horizontal beam size $\sigma_m$ can be obtained from a Gaussian fit of the profile. Fig. 10 shows the measured $\sigma_m$ as a function of the iris diameter. By deconvolution $\sigma_r^2 = \sigma_m^2 - \sigma_x^2$, the optical resolution was obtained. Because the acceptance was limited by the projected horizontal width of the Be mirror (22/√2=15.6mm), a constant resolution 325±15 μm was observed when the iris diameter was greater than 15.6 mm. When the iris diameter was less than 7 mm, the resolution was dominated by the diffraction (0.61$\lambda/\theta_h$, where $\theta_h$ is the horizontal acceptance), and the measured resolution agreed reasonably well with the theoretical value [4, 6]. Although the theoretical calculation predicts a minimum resolution when the iris diameter is near 11 mm, it consistently underestimated the resolution. Given the experimental resolution, we can measure larger horizontal beam sizes greater than 500 μm which are well beyond the range of the interferometer. An example will be discussed in the following section.

*3.3 Transverse beam size measurement in an undulator lattice*

Recently, Cornell High Energy Synchrotron Source (CHESS) developed a compact in-vacuum undulator [31]. The undulator was installed in CESR for a test run. A 5.3 GeV undulator lattice has a design horizontal emittance of 54.2 nm. As listed in Table 2, the expected horizontal beam size at the vBSM source point is 847 µm, too large to be measured with the interferometer with our slit spacings. Therefore, the direct imaging method was used to determine the horizontal beam size.

*3.3.1 Horizontal beam size*

Fig. 11a shows a typical direct image of the beam profile obtained in the undulator lattice with a beam current of 0.9 mA. The integrated horizontal intensity profile is Gaussian with width corresponding to a horizontal beam size of $\sigma_m$=946±3 µm (Fig. 11b). The averaged horizontal beam size measured in one minute is 943±10 µm (Fig. 11c). After subtracting the resolution $\sigma_r$=325 µm in quadrature, the corrected horizontal beam size is $\sigma_x$=885±15 µm. Given the theoretical energy spread and the horizontal dispersion at the vBSM source point (Table 2), we extract horizontal emittance: $\varepsilon_x$=60.7±2.7 nm, which is slightly larger than the theoretical design value of 54.2 nm.

*3.3.2. Vertical beam size*

In order to measure the vertical beam size in the undulator lattice we replaced the triplet of vertically aligned double slits with a triplet of horizontally aligned slits. The vertical beam size is approximately an order of magnitude smaller than the horizontal (70 µm versus 885 µm), so separation of the three pairs of horizontally aligned slits is increased with respect to the vertical slits to *D*=3.0, 5.0, and 8.0 mm. Other optical elements remain at the same positions. The interferometer is then employed to measure the vertical beam size.

With a 0.9 mA beam in CESR, we recorded the interference patterns using all three pairs of slits as shown in Fig. 12a, 12b and 12c, respectively. Their intensity fringes are shown in Fig. 12d and the visibilities $\gamma$ are obtained from fitting these fringes (Fig. 12e). As expected, slits with closer separation yield higher visibility. By fitting the visibility as a function of the slit separation (Eq. 3), we acquire a vertical beam size of $\sigma_y$=67±1 µm, corresponding to a vertical emittance of 0.25±0.01 nm and an emittance coupling of 0.4%.

By moving the double slits out of SR light path and rotating the polarizer by 90°, we used the CCD camera to image the π-polarized component of SR light. Fig. 13a shows an example image of the π-polarized component, which has a similar structure to the simulated intensity distribution shown in Fig. 3a. The integrated horizontal intensity profile (Fig. 13c) yields a horizontal beam size of 932±3 µm, consistent with that measured with the σ-polarized component (Fig. 11a).

From the vertical integrated intensity profile (Fig. 13b), we obtain the valley-to-peak intensity ratio of 0.431. Comparison with the simulated valley-to-peak ratio as a function of $\sigma_y$ (Fig. 2c) yields a vertical beam size of 71.8 µm, in reasonable agreement with the beam size measured using the interferometer. The averaged vertical beam size measured over a period of two minutes is 73.4±4.1 µm, greater than the averaged vertical beam size from the interferometer 70.1±2.0 µm as shown in Fig. 13d. An advantage of using the π-polarization method is that both horizontal and vertical beam size can be measured simultaneously.

*3.3.3 Measurement error*

As discussed in section *3.2.1*, the curvature of the trajectory and the depth of field contribute a systematic error to the horizontal beam size measurement with the interferometer. Due to finite depth of field the observed vertical beam size is effectively an average over some longitudinal distance [13]:

$$\gamma_v = \iint f(s,y)\exp(-i\frac{2\pi D}{\lambda L}y)dsdy = \int \bar{f}(y)\exp(-i\frac{2\pi D}{\lambda L}y)dy \qquad (11)$$

where *f(s,y)* denotes the vertical beam profile at longitudinal position *s* and *f(y)* is the averaged vertical beam profile over some longitudinal distance. In an accelerator, the vertical beam size varies with $\sqrt{\beta_y}$. In the undulator lattice, the change in $\beta$ is ±4% over a distance of ±0.18 m (depth of field) with respect to the vBSM source point. The expected variation in vertical beam size over that distance is ±2%, which is small. Furthermore, since the vertical beam size changes approximately linearly with longitudinal position near the source point, the average vertical beam size over this longitudinal distance is very close to the vertical beam size at the source point. Since the observed beam size is essentially the average beam size, the error due to the depth of field (0.35m) is negligible.

Another source of error in the interferometric measurement of vertical beam size is beam jitter. Assuming the vertical beam jitter is also 50 μm (comparable to the horizontal) and the double slits have separation *D*=5.0 mm, we obtain from Eq. 9 and 10, a phase shift of $\Delta\psi$=0.52 and an average visibility of $\bar{\gamma}$ =0.955$\gamma$. This 4.5% decrease in visibility corresponds to a 6 μm (8.6%) increase of the measured beam size, which is a nonnegligible effect.

In addition, the beam jitter also significantly affects the measured vertical beam size using the $\pi$-polarization method. Although short exposure time less than 2 ms was used to acquire the CCD images, so as to eliminate any contribution from systematic vibration with frequency less than 500 Hz, beam jitter at higher frequencies cannot be avoided. A single CCD camera image is an integrated image of multiple randomly-jittered beam profiles over a period of about 2 ms. The random beam jitter smears the visibility of the $\pi$-polarized intensity distribution of SR. Since the measured vertical beam size relies on the averaged intensity profile of the acquired image, the beam jitter amplitude will directly affect the measured beam size. The larger the beam jitter, the greater the valley-to-peak intensity ratio. As shown in Fig. 13d, we observe a larger spread in the measured vertical beam size by the $\pi$-polarization method than the interferometric method, indicating that the $\pi$-polarization technique for measuring beam size is more sensitive to beam jitter than the measurement with the interferometer.

In order to eliminate sensitivity to beam jitter we plan to employ a gated camera [32] or photomultiplier array with single bunch turn-by-turn capability (~ 2 ns resolution).

*3.4 Bunch length measurement*

As shown in Fig. 1, a beam splitter on the optical table directs a fraction of the visible SR into a streak camera setup [3], demonstrating the concept for simultaneous measurement of bunch length and transverse sizes. Indeed, we have successfully measured the horizontal beam size with the interferometer and the bunch length with streak camera simultaneously during intra-beam scattering studies [29]. By employing the $\pi$-polarization method, the vBSM can in principle provide a measure of the beam size in three dimensions simultaneously. An alternative approach is to replace the double slits with quadrate slits to construct a 2-dimensional interferometer [21, 22]. However, the precision of measuring the transverse beam sizes may not be as high as using a single interferometer [14].

**4. Conclusion**

A visible beam profile monitor was designed and installed in CESR, and was used to measure transverse beam sizes via both direct imaging and interferometric methods. The interferometer was implemented to measure the horizontal and vertical beam sizes over a wide range with good accuracy. The π-polarization method for measuring the vertical beam size was explored in an undulator lattice, and showed good agreement with the measurement made with the interferometer. Employing the π-polarization method, we demonstrated the capability to measure the beam sizes in all three dimensions simultaneously. The vBSM was successfully used to measure the horizontal beam size and the bunch length for IBS studies. Detailed analysis of systematic effects has also been presented.

**Acknowledgment**

The authors thank J.W. Flanagan, M. Billing, J. Shanks, M. Ehrlichman, and W. Hartung for valuable discussions. This research was supported by the National Science Foundation and Department of Energy under contract numbers PHY-0734867, PHY-1002467, PHYS-1068662, DMR-0225180, and DE-FC02-08ER41538, DE-SC0006505.

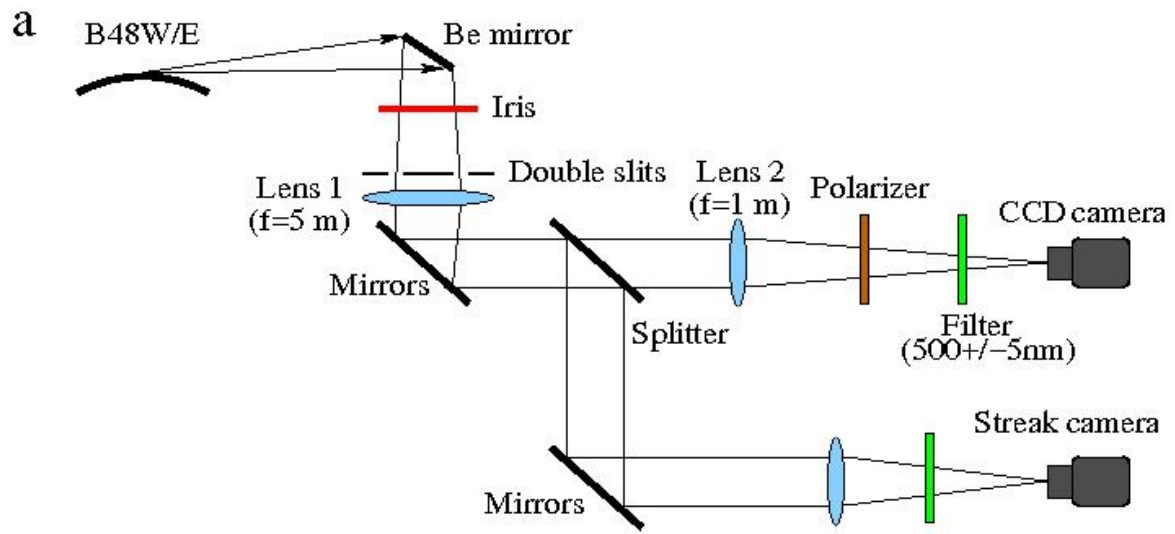

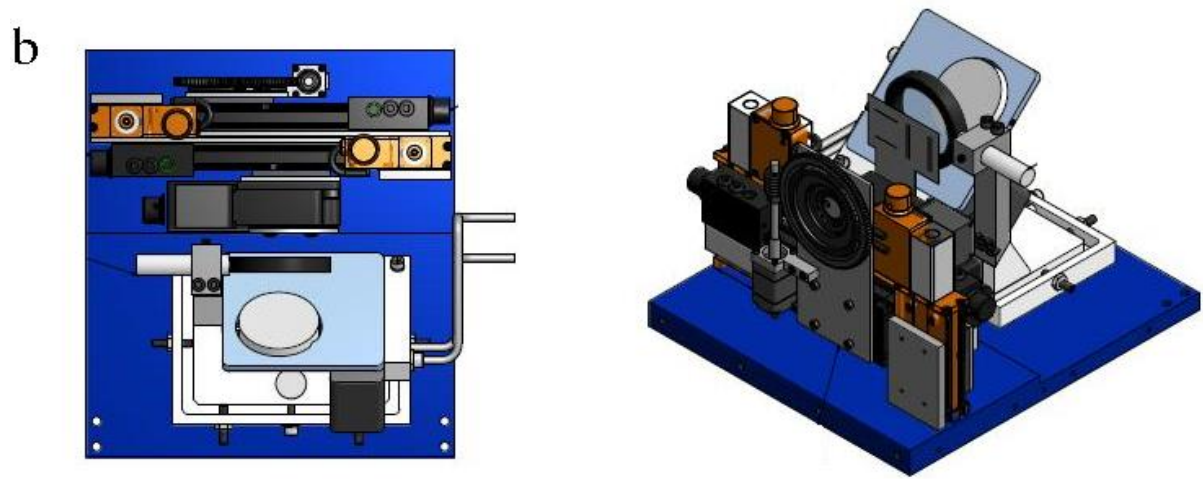

Fig. 1. (a) The schematic setup of visible beam size monitor. (b) The optics box containing the optical elements: an iris with adjustable aperture, a double slit set, a lens, and a mirror.

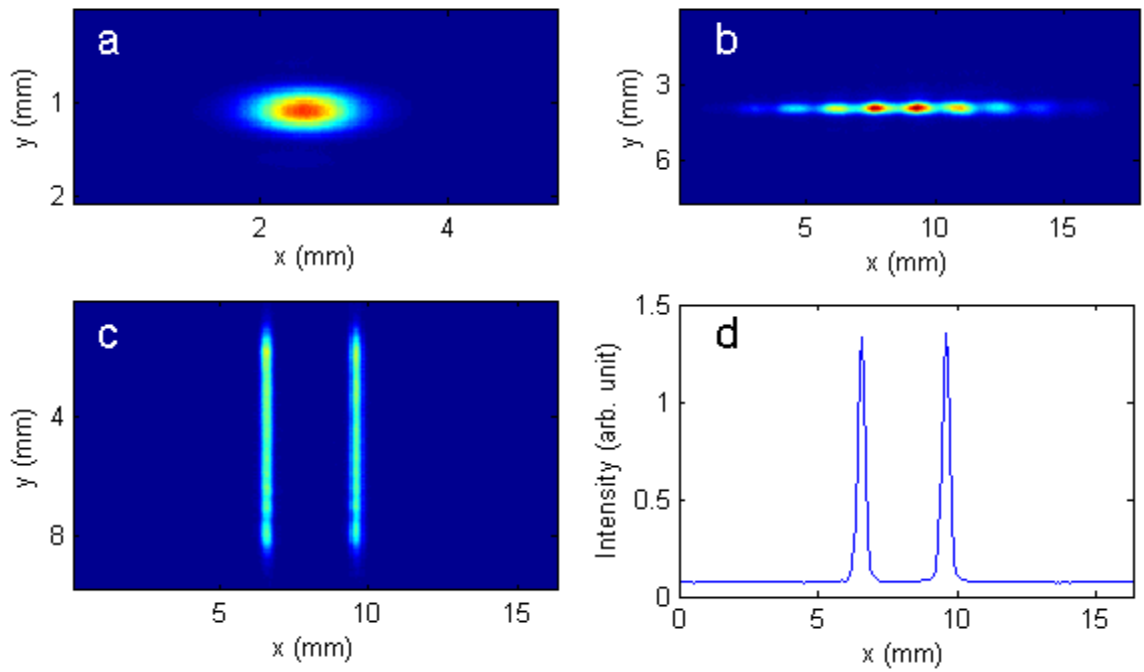

Fig. 2. Typical CCD images recorded at two different image planes. (a) direct image and (b) interferometer image are at the image plane of the source point and (c) is at the image plane of the double slits ($D$=3.0 mm). (d) The integrated horizontal intensity profile of (c).

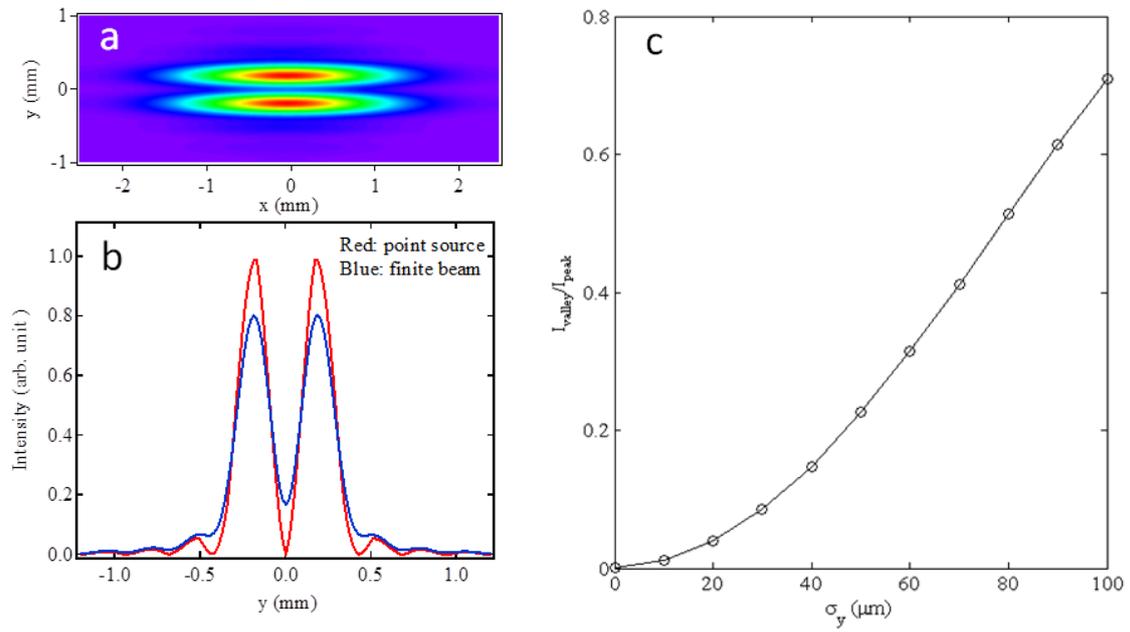

Fig. 3. Simulation using SRW code. (a) The intensity distribution of vertically polarized SR light from a soft bend (140.6 m bend radius) in CESR assuming a finite beam ($\sigma_h$=800 μm and $\sigma_v$=50 μm) at 5.3 Gev. (b) The intensity profile at x=0. The red line is from a point source while the blue line is from the finite beam. (c) The valley-to-peak intensity ratio as a function of vertical beam size.

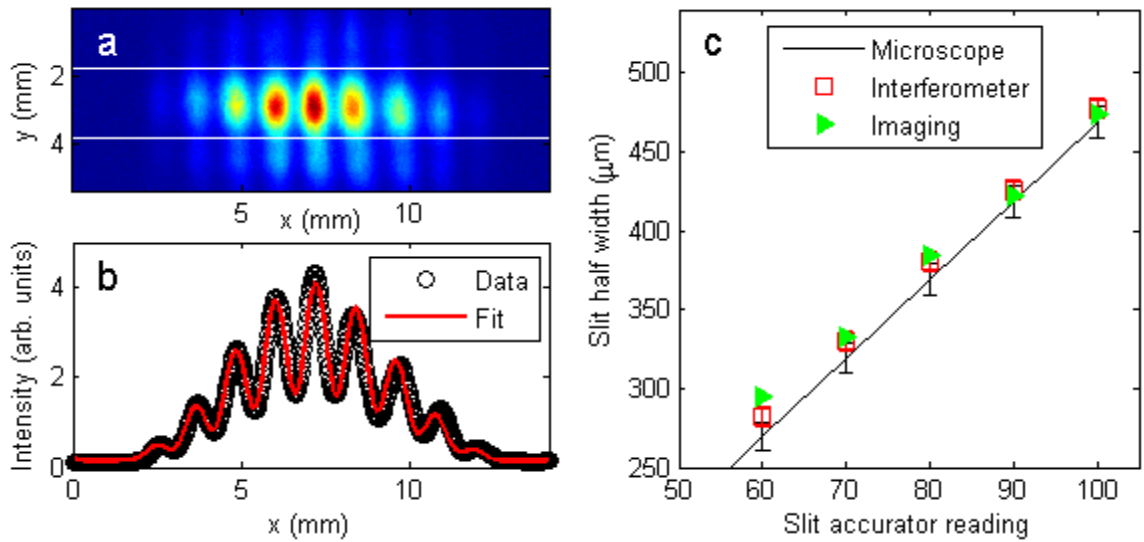

Fig. 4. (a) An interference pattern from a filament source behind an adjustable single slit with a half width of 370 μm using $D$=2.5 mm double slits. (b) The horizontal intensity profile integrated between two white lines in (a) and the best fit using Eq. 1. (c) The measured half width of the single slit from three different methods.

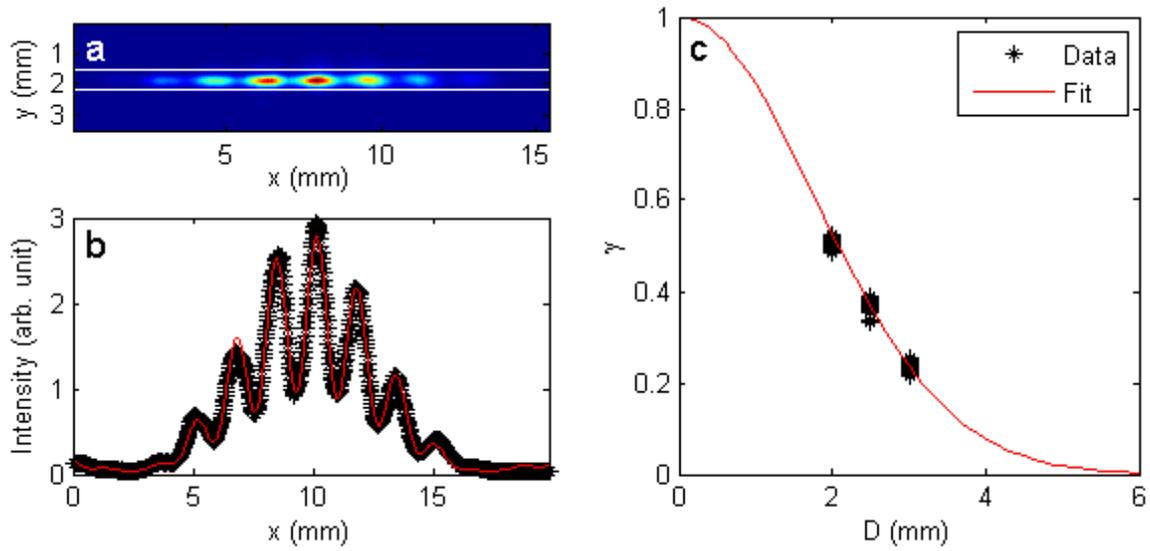

Fig. 5. (a) A typical interference pattern of SR using a *D*=2.0 mm double slits. (b) The horizontal intensity profile integrated between two white lines in (a) and the best fit using Eq. 1. (c) The visibility measured using three different sets of double slits (*D*=2.0, 2.5, and 3.0 mm). The red line is the best fit using Eq. 3.

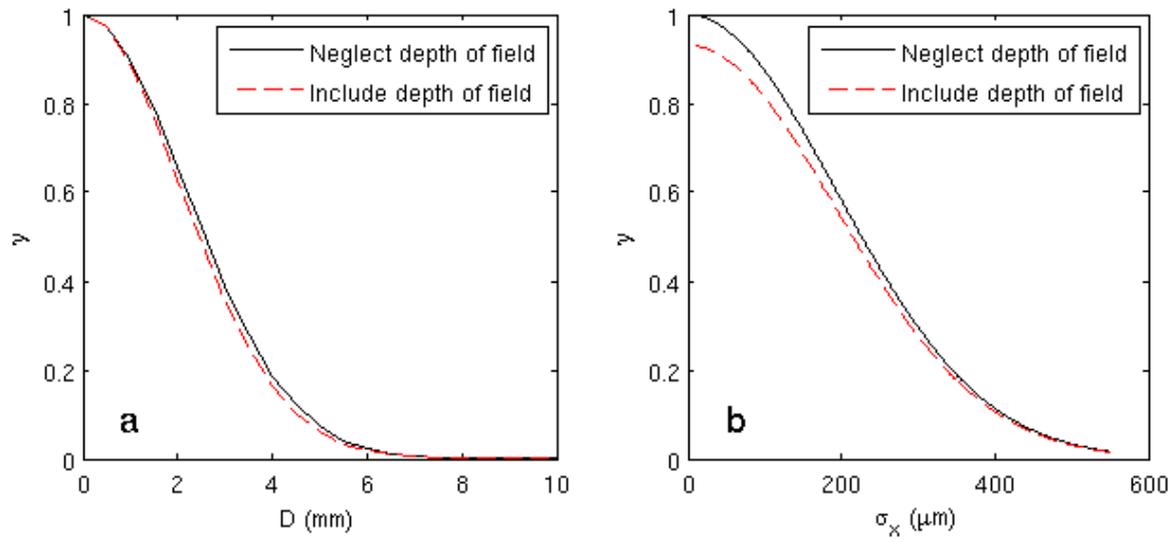

Fig. 6. Simulated visibility as a function of (a) *D* and (b) $\sigma_x$ ignoring (Eq. 2) and including depth of field (Eq. 8).

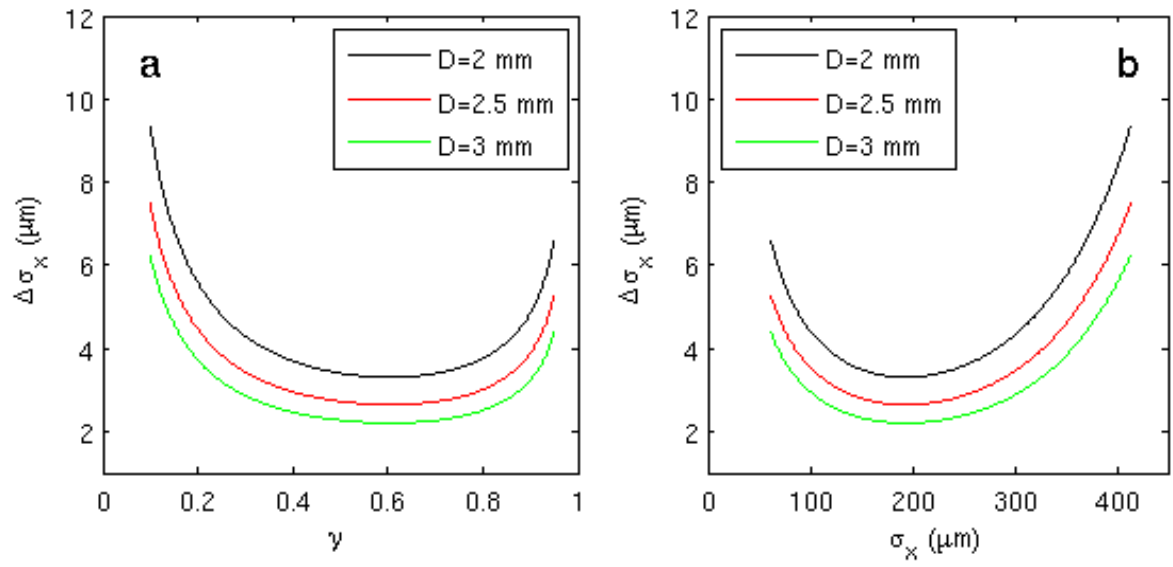

Fig. 7. Beam size error as a function of (a) $\gamma$ and (b) $\sigma_x$ for three different double slits.

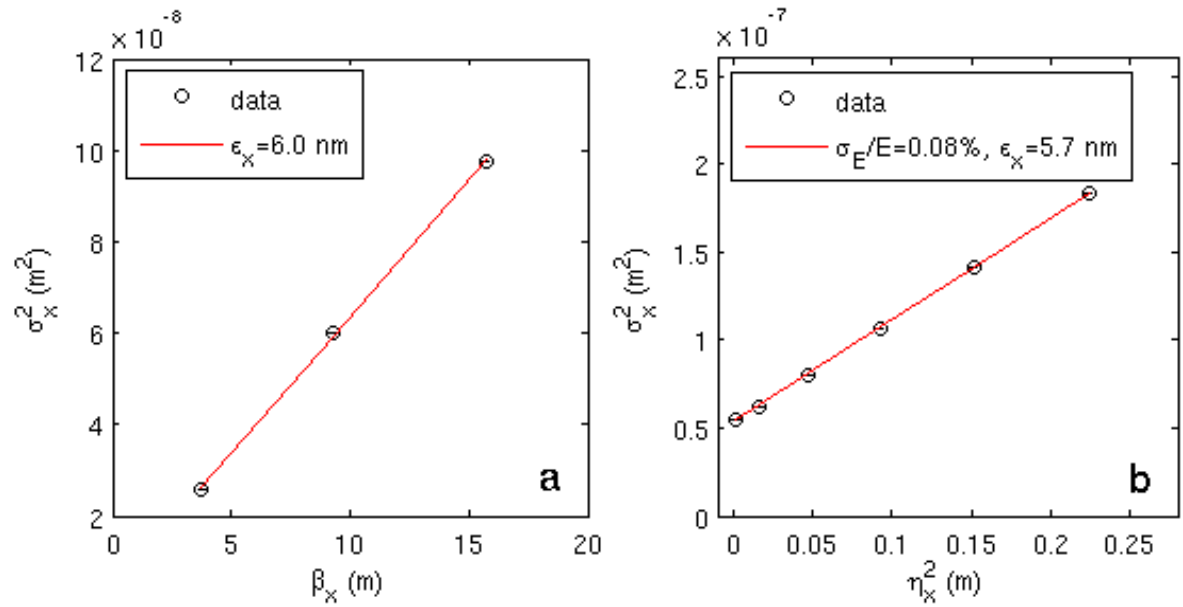

Fig. 8. The corrected horizontal beam size as a function of (a) $\beta_x$ and (b) $\eta_x$. Symbols are the data and the line is the best linear fits to the data.

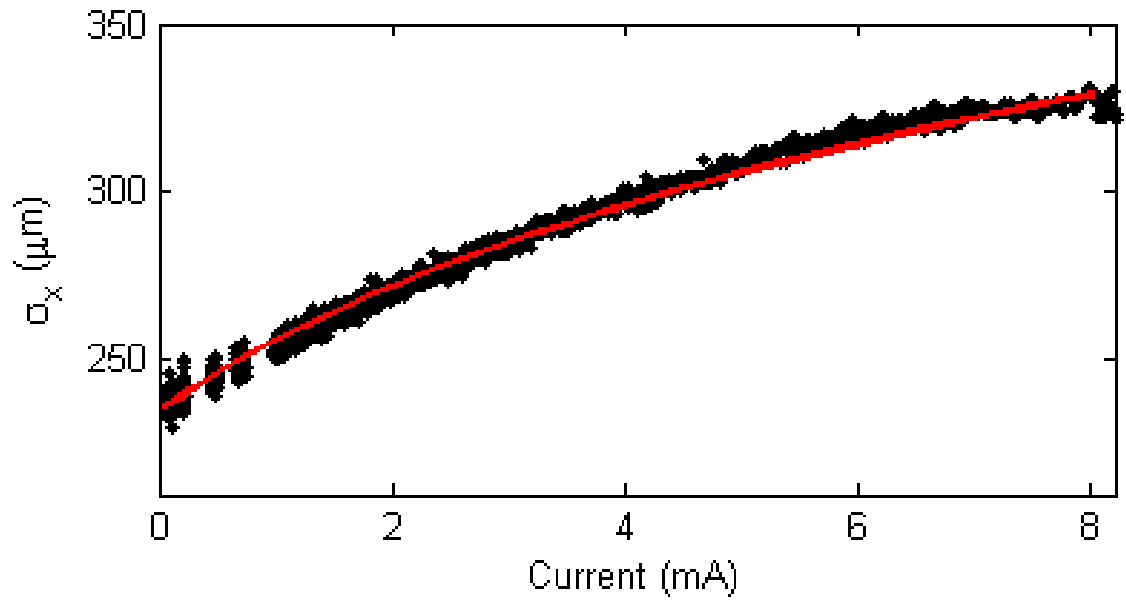

Fig. 9. The corrected horizontal beam size as a function of beam current for an IBS study. Symbols are the data and the red line is the best fit from IBS simulation.

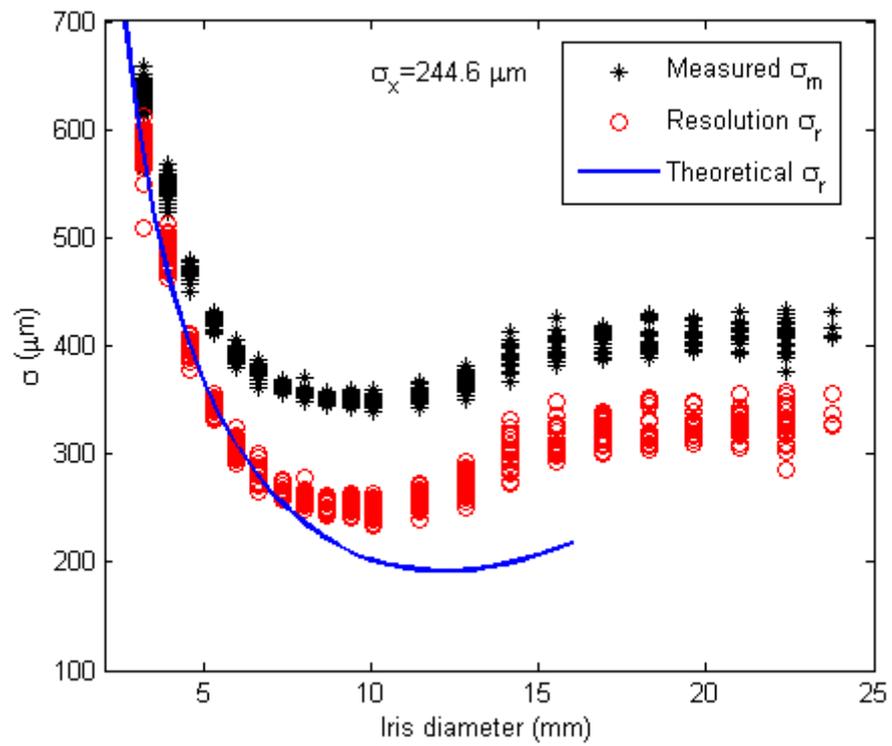

Fig. 10. The measured horizontal beam size (black stars), the calculated resolution (red circles), and the theoretical resolution (blue line) as a function of the iris diameter.

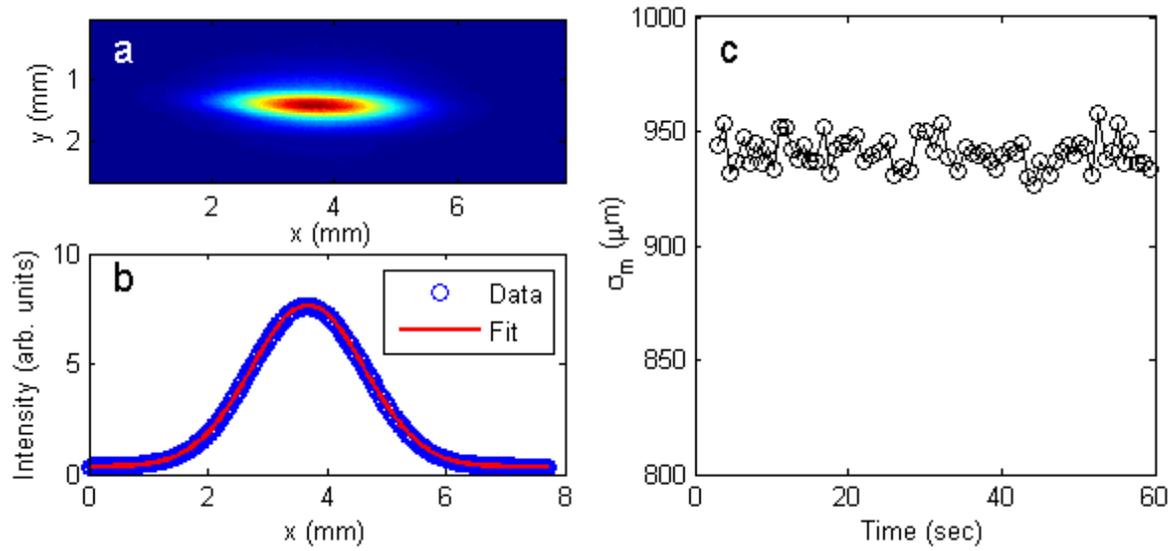

Fig. 11. The horizontal beam size measurement using the direct imaging method in an undulator lattice. (a) The measured intensity distribution of σ-polarized SR at the image plane of the source point. (b) The horizontal intensity profiles integrated over all the vertical pixels in image (a) and the best Gaussian fit. (c) The horizontal beam size measured over a minute time interval.

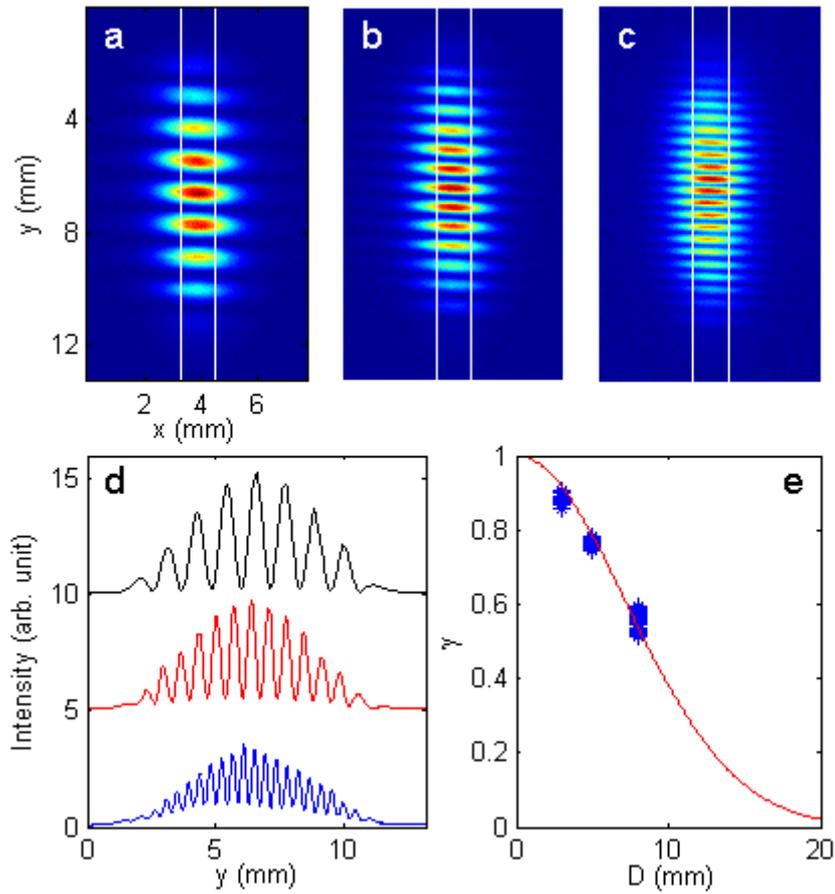

Fig. 12. A vertical beam size measurement using interferometer in an undulator lattice. Three different double slits, (a) D=3.0, (b) 5.0, (c) 8.0mm, were used to acquire the interference fringes. (d) From top to bottom, the vertical intensity profiles integrated between two white lines in (a), (b), and (c), respectively. (e) The measured visibility as a function of slit separation. The symbols are the data and the line is the best fit using Eq. 3.

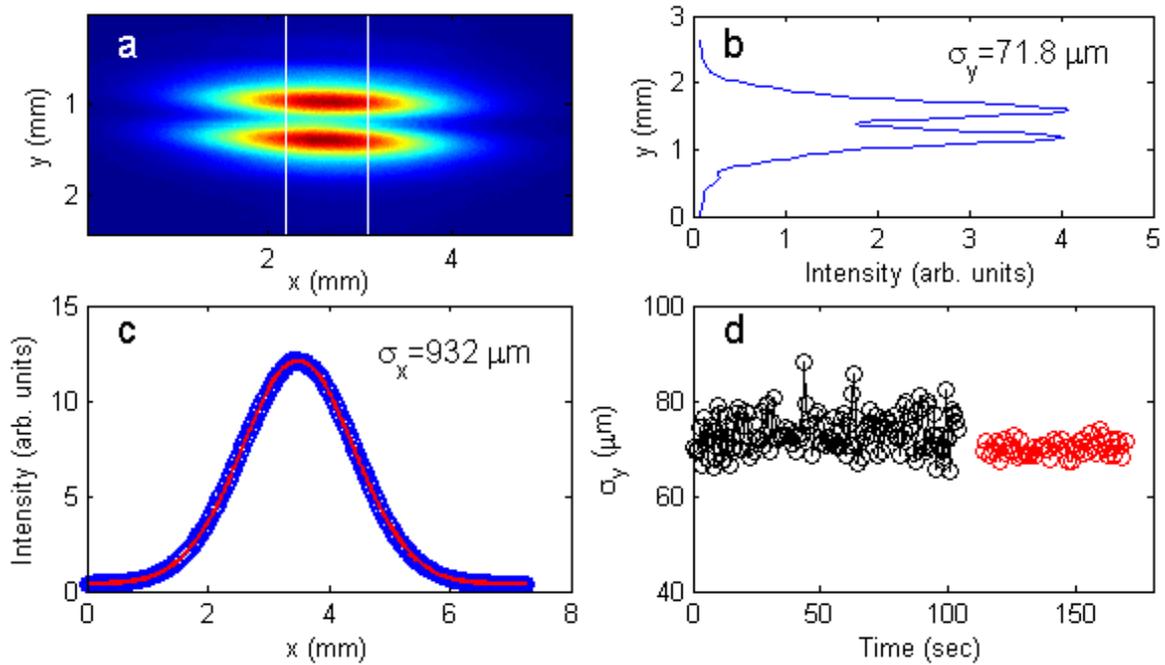

Fig. 13. A transverse beam size measurement using π-polarization method. (a) The measured intensity distribution of π-polarized SR at the image plane of the source point. (b) The vertical intensity profile integrated between two lines in (a). (c) The horizontal intensity profile integrated over all the vertical pixels in image (a). (d) The vertical beam size measured in real time using π-polarization method (black circles) and the interferometer with a D=5.0mm double slits (red circles).